\documentclass[aps,prd,twocolumn,nofootinbib,notitlepage]{revtex4-1}
\usepackage{amsmath,amssymb,amstext,graphicx,color,xfrac,braket,accents,tabulary,booktabs,xcolor,bbold,hyperref}
\begin{document}

\baselineskip=13pt

\vspace{0.5cm}
\thispagestyle{empty}

\newcommand{\be}{\begin{equation}}
\newcommand{\ee}{\end{equation}}
\newcommand{\bea}{\begin{eqnarray}}
\newcommand{\eea}{\end{eqnarray}}
\def\lag{{\mathcal L}}
\def\E{{\bf E}}
\def\B{{\bf B}}
\def\F{\widehat{\bf B}}
\def\G{\widehat{\bf E}}
\def\A{{\bf A}}
\def\x{{\bf x}}
\def\k{{\bf k}}
\def\u{{\bf u}}
\def\v{{\bf v}}
\def\J{{\bf J}}
\def\t{{\hat t}}
\def\mn{{\mu\nu}}
\def\wt#1{\widetilde{#1}}
\def\g{{g_\varphi}}
\newcommand{\draftnote}[1]{\textbf{\color{red}[#1]}}

\vspace*{-0.4cm}
\hfill \today

\title{Axions and Cosmic Magnetic Fields}
\author{George B. Field$^a$ and Sean M. Carroll$^b$}
\affiliation{{}$^a$Harvard-Smithsonian Center for Astrophysics, Cambridge, MA 02138; gravturb@yahoo.com\\
{}$^b$Johns Hopkins University, Baltimore, MD 21218; seancarroll@gmail.com}

\begin{abstract}
We argue that if axions are the dark matter, their coupling to electromagnetism $\varphi F_\mn \widetilde{F}^\mn$ results in exponential growth of a helical magnetic field when the axion field first rolls down its potential.
After an inverse cascade, the relevant length scales today are $\sim 10^1$-$10^2$\,kpc, of astrophysical interest.
Our mechanism for allowing the field to grow relies on a nuance of MHD.
Faraday's Law says that an electric field is needed to create a magnetic field. 
Previous authors relied on conventional Ohm's law to calculate $\E$, but the resistivity is negligible and therefore they assume $\E$ is as well.
We use a modified Ohm's Law that includes the effects of self induction in limiting the current driven by a given $\E$, which allows a magnetic field to grow.
\end{abstract}

\maketitle

\section{Introduction}

Magnetic fields are observed within galaxies and within galaxy clusters, but so far, not reliably so outside of galaxy clusters.
However, Dolag et al \cite{2002A&A...387..383D} propose that the fields observed within clusters were created before the clusters themselves, and accreted into them as they formed.
It is possible that such fields are of primordial origin.
This possibility is attractive, because it could open new ways to learn about the early universe.
For reviews see \cite{widrow2002origin,subramanian2016origin,Vachaspati:2020blt}.

One promising candidate for the epoch of magnetogenesis is the QCD phase transition at 300~MeV, which takes place at $t \sim 10^{-5}$\,s.
However, any fields produced at that era would be coherent over at most the Hubble length at that time, $\lesssim 10^{5}$\,cm.
If the fields were frozen into the comoving plasma, even taking into account the Hubble expansion of $\sim 10^{12}$, the scales today would be measured in AU, too small to be observed.

Field and Carroll \cite{Field:1998hi} (hereafter FC) noted that if the field is helical, nonlinear interactions cause magnetic energy to move to large scales \cite{1997PhRvD..56.6146C,1999PhRvD..59f3008S}.
FC used previous simulations \cite{pouquet_frisch_leorat_1976} to show that an inverse cascade would cause an increase in scale by a factor of $10^8$ over and above that due to the Hubble expansion, so 1\,AU becomes 1\,kpc.

However, this scenario requires that the field is helical.
The generation of magnetic helicity 
\be
  H = \int \A \cdot \B \, d^3x
  \label{helicity}
\ee 
is governed by
\be
  \partial_t H = -2 \int \E \cdot \B\,d^3x ,
  \label{dtH}
\ee
in units where $\hbar=c=k=1$, which we assume henceforth.
The standard MHD expression for $\E$ is
\be
  \E = \eta \J + \v\times\B,
  \label{standardE}
\ee
where $\J$ is the current and $\eta$ is the resistivity.
In our case, $\eta$ is minuscule \cite{baym1997electrical}, so mechanisms based upon finite $\eta$ don't work, leaving only $\v\times\B$.
But that term is perpendicular to $\B$, and therefore doesn't contribute to the helicity, following (\ref{dtH}).

It follows that $\partial_t H\neq 0$ is contrary to conventional MHD when the resistivity is negligible.
We have therefore studied possible modifications to (\ref{standardE}) and have found one:
\be
   \E = \eta \J + \v\times\B + D\partial_t \B,
  \label{newE}
\ee
where $D$ is the size of the region, for us the Hubble radius at the QCD transition.
The new term is due to self induction and it is responsible for the term $L\frac{dI}{dT}$ in circuit theory, as explained in Section~\ref{induction}.

The rest of this paper shows how (\ref{newE}) can lead to successful production of helical magnetic fields in the presence of an evolving pseudoscalar axion field $\varphi$ (taken to be dimensionless for the moment).
The axion couples to electromagnetism through
\be
  \lag = -\frac{\g}{4}\varphi F_\mn \wt{F}^\mn.
  \label{phiFFdual}
\ee
Here $\wt{F}^\mn$ is the dual tensor
\be
  \wt{F}^\mn = \frac{1}{2}\epsilon^{\mu\nu\rho\sigma}F_{\rho\sigma}, \label{fdual}
\ee
where $\epsilon^{\mu\nu\rho\sigma}=|g|^{-1/2}e^{\mu\nu\rho\sigma}$, $e^{\mu\nu\rho\sigma}$ is the alternating symbol with $e^{01234} = -1$, $g = -\mathrm{det}(g_\mn)$, and $\g$ is a dimensionless coupling constant.
Taking a homogeneous pseudoscalar $\nabla\varphi = 0 $ for simplicity, the modified Maxwell equations are
\bea
  \partial_t\E + 4\pi \J -  \nabla\times\B + \g\dot\varphi \B &=& 0,\\
  \partial_t\B +  \nabla \times\E &=& 0.
\eea
The challenge is to find solutions to these equations that yield $\E\cdot\B \neq 0$.
This idea has been pursued in the context of inflation, but generally the produced fields are too small \cite{1988PhRvD..37.2743T,1992PhRvD..46.5346G}.
FC tried one based on $\E = \eta\J$, which yielded small values of $\A\cdot\B$ because $\eta$ is so small.
Campanelli and Giannotti \cite{Campanelli:2005ye} and Miniati et al. \cite{miniati} identified $\varphi$ with the axion and also assumed that $\E = \eta\J$, with the same disappointing results.
For related approaches see \cite{1996PhRvD..54.1291B,1997PhRvL..79.1193J,1997PhLB..398..321O,Brandenburg:2018ptt,PhysRevD.102.103528,Mtchedlidze_2022}.

In this paper we assume that $\eta = 0$ and $\v=0$, so $\E=0$ according to (\ref{standardE}), but can be non-zero according to (\ref{newE}).
It is this new feature that allows for the generation of a significant magnetic field.

\section{Pseudoscalar Evolution} \label{pseudoscalar}

Our mechanism for producing helicity is based on introducing a pseudoscalar $\varphi$ that couples as in (\ref{phiFFdual}).
The QCD axion, proposed to solve the Strong-CP problem \cite{1977PhRvL..38.1440P,PhysRevLett.40.223,PhysRevLett.40.279} and also a promising dark matter candidate \cite{Preskill:1982cy,Abbott:1982af,Dine:1982ah,Duffy:2009ig}, is one such hypothetical field, although  other models have been considered.
Note that this interaction has no effect on the dynamics in any domain where $\varphi$ is constant, as that would only multiply $F_{\alpha \beta} \tilde{F}^{\alpha \beta}$ by a constant, and this term is a total derivative that doesn't contribute to the equations of motion. 
Therefore we anticipate that helicity is generated only in regions of spacetime where $\varphi$ is changing. 
This will be the case, for example, at the QCD phase transition, where the axion potential becomes relevant, and $\varphi$ evolves from an initial value $\varphi_0$ down to zero.
In this section we solve the pseudoscalar equations of motion numerically, then in the next section we will introduce simple approximations that will allow us to understand dependence on the relevant parameters.

Since our ultimate goal is generating magnetic fields rather than something like inflation, we will assume for simplicity that the pseudoscalar itself provides a negligible fraction of the energy density of the universe at the QCD scale (though it will account for dark matter in the matter-dominated era), and later check that our assumption is self-consistent.
We will likewise start by assuming that the generated magnetic field does not back-react in any important way on $\varphi$ or the scale factor.

We consider a flat Robertson-Walker cosmology, so the metric is
\be
  ds^2 = -dt^2 + R^2(t)d\vec{x}^2.
  \label{rwmetric}
\ee
The Friedmann equation in a flat radiation-dominated universe is
\be
  H^2 = \frac{8\pi G}{3} \rho(T),
  \label{feq}
\ee
where $T$ is the temperature and $\rho$ is the energy density,
\be
  \rho(T) = \frac{\pi^2}{30} g_*(T) T^4,
\ee
where $g_*$ is the effective number of relativistic degrees of freedom \cite{Kolb:1990vq}.
We will be concerned with times just before the QCD phase transition, with $T\sim 1$~GeV, and for convenience will approximate $g_*$ as the constant value $g_*=80$.
Then $T\propto R^{-1}$ and the solution to (\ref{feq}) is $R \propto t^{1/2}$, with
\be
   \frac{T}{\mathrm{MeV}} = \frac{1.6}{ g_*^{1/4} }\left(\frac{t}{\mathrm{sec}}\right)^{-1/2}
   = 0.53 \left(\frac{t}{\mathrm{sec}}\right)^{-1/2}
\ee
and
\be
  H = \frac{1}{2}t^{-1}.
\ee
It will be convenient to measure time in nanoseconds ($10^{-9}\,$sec), so we have
\be
   \frac{T}{\mathrm{GeV}} 
   = 1.7 \times 10^1 \left(\frac{t}{\mathrm{ns}}\right)^{-1/2}.
   \label{temperature}
\ee
The QCD phase transition $T_\mathrm{QCD}=300\,$MeV  is at $t (\Lambda_\mathrm{QCD}) = 3.2\times 10^{3}$\,ns.
For definiteness we set the zero-temperature axion mass to be $m_0 = 10^{-5}$\,eV$= (6.6 \times 10^{-2}\, \mathrm{ns})^{-1}$,
using $1\,\mathrm{eV} = 1.5\times 10^{15} \, \mathrm{sec}^{-1}$.
This is consistent with values that allow the axion to be a candidate for cold dark matter \cite{Workman:2022ynf}.

Let us consider for the moment a canonically-normalized scalar field $\phi$, before returning to our dimensionless variable $\varphi = \phi/f$, where $f$ is the Peccei-Quinn scale or axion decay constant.
The action is
\be
  S_\phi = \int d^4x \sqrt{-g} \left[-\frac{1}{2}g^{\mu\nu}\partial_\mu\phi\partial_\nu\phi - V(\phi)\right].
\ee
The scalar equation of motion is 
\be
  \ddot\phi + 3H \dot\phi + \frac{dV}{d\phi} = 0.
\ee
The axion potential $V(\phi)$ is temperature-dependent, as it comes to life during the QCD phase transition.
Following Marsh \cite{Marsh:2017hbv}, we can write it as
\be
  V(\phi) = m_a^2(T) f^2[1-\cos(\phi/f)], 
\ee
where for simplicity we assume there is a unique vacuum.

The function $m_a(T)$ is the temperature-dependent axion mass, given by the second derivative of the potential at its minimum.
The axion potential is flat at high temperatures, turns on as the temperature approaches the QCD scale, and reaches a fixed form thereafter \cite{di_Cortona_2016,Kim_2018}.
For high temperatures it takes the form
\be
  m_a(T> \Lambda_\mathrm{QCD}) = m_0\left(\frac{T}{\Lambda_\mathrm{QCD}}\right)^{-4},
\ee
where $m_0$ is the zero-temperature mass, and $\Lambda_\mathrm{QCD} = 300$~MeV.
We will use the overall functional form
\be
  m_a(T) = \frac{m_0}{1+(T/\Lambda_\mathrm{QCD})^4}.
\ee
We can plot the axion mass in eV as a function of both temperature (in GeV) and time (in ns).
\begin{figure}[h]
  \begin{center}
    \includegraphics[width=0.45\textwidth]{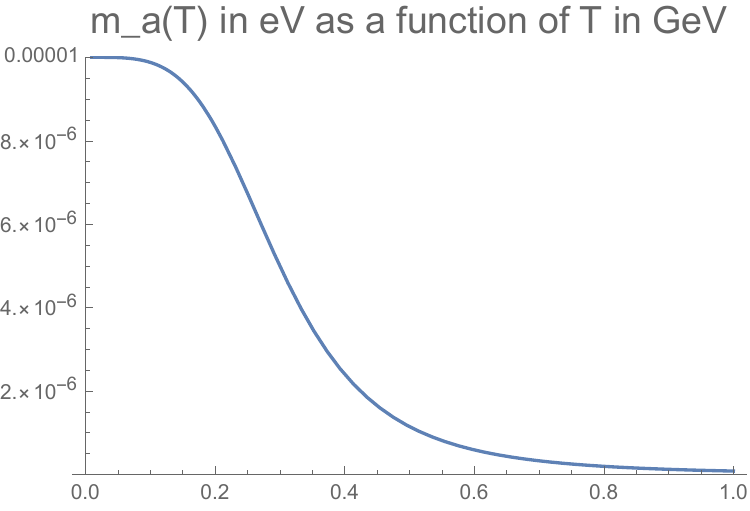}    
    \includegraphics[width=0.45\textwidth]{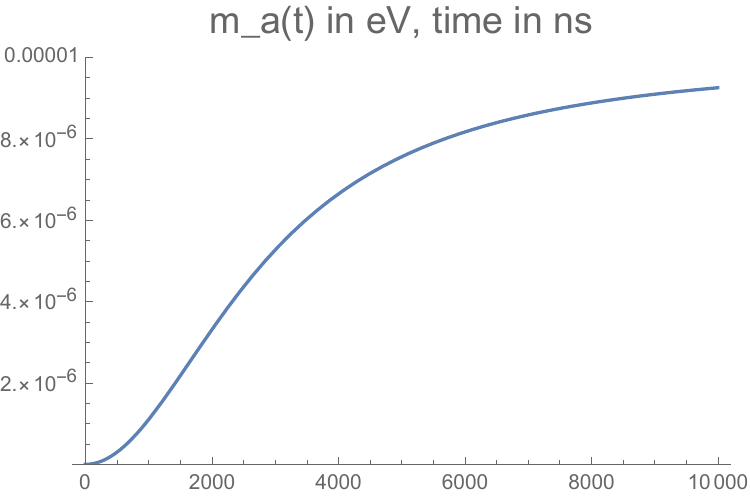}
  \end{center}
\end{figure}
In these plots we have taken $m_0 = 10^{-5}$\,eV and $f=10^{12}$\,GeV.

The scalar equation to be solved is thus
\be
  \ddot\phi + \frac{3}{2}t^{-1}\dot \phi + m_a^2(T)f\sin(\phi/f) = 0.
\ee
Now we can convert back to our dimensionless variable $\varphi = \phi/f$, so this becomes
\be
  \ddot\varphi + \frac{3}{2}t^{-1}\dot \varphi + m_a^2(T)\sin(\varphi)=0. 
  \label{varphieq}
\ee

\begin{figure}[h]
  \begin{center}
    \includegraphics[width=0.45\textwidth]{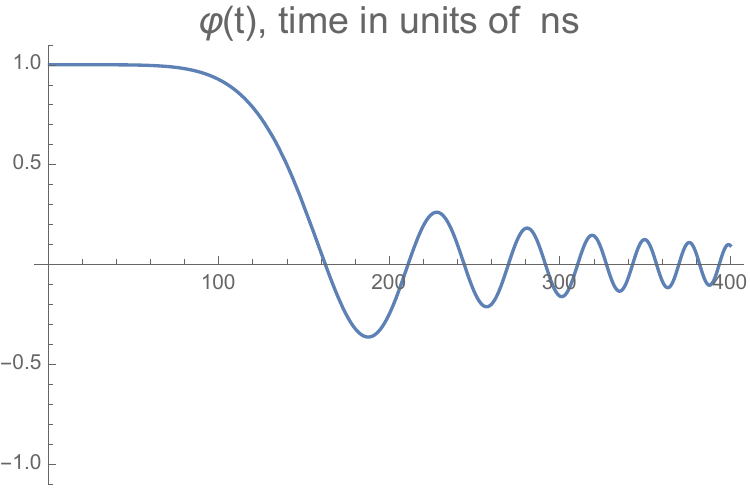}
  \end{center}
\end{figure}

Solving numerically, we see $\varphi$ is initially stuck at its starting value, then starts rolling near $t = 10^{2}$\,ns.
(The dimensionless variable $\varphi= \phi/f$ takes on values between 0 and $\pi$, the angle in the brim of a Mexican hat potential, so we have chosen $\varphi_0 = 1$ for this numerical example.)
Note that this is actually somewhat earlier than $t_\mathrm{QCD}= 3.2\times 10^{3}$\,ns.
That's because the field starts rolling when $m_a \approx H$, which happens relatively quickly once the potential begins to develop at all.

\section{The Role of Induction}\label{induction}

In the absence of electrical charges, and therefore zero conductivity, the coupling $\varphi F\widetilde F$ can lead to exponential growth in the magnetic field \cite{1990PhRvD..41.1231C,1992PhRvD..46.5346G}.
This makes this physical process seem promising for generating large primordial fields.
Typically, however, we imagine that the early universe is full of charged particles, and therefore the conductivity is high.

For very large conductivity $\sigma\rightarrow \infty$, conventional wisdom in MHD holds that no electric field can be generated.
Here we suggest that this does not necessarily hold.
Although more careful analysis is needed to rigorously establish our conjecture, here we explore its implications for magnetogenesis.

Let $\E'$ represent the field as measured in the rest frame of the plasma.
Then 
\be
  \E' = \sigma^{-1}\J = \eta \J \rightarrow 0.
\ee
But in the lab frame, with $\v$ representing the plasma velocity, we have
\be
  \E = \E' + \v\times \B.
\ee
For clarity, consider the case $\v = 0$. Then if $\sigma \rightarrow 0$, we have
\be
  \E = \E' \rightarrow 0.
\ee
Below we will show that even this is wrong.

It is all because
\be
  \partial_t\B = -\nabla\times\E,
\ee
Maxwell's homogeneous equation, which is equivalent to Faraday's Law
\be
  \oint \E\cdot d{\bf S} = \partial_t\Phi,
  \label{magneticflux}
\ee
where $\Phi = \oint \B\cdot d{\bf S}$.
It follows that if we are interested in $\partial_t\B$, then $\E$ must be finite.

This is recognized in circuit theory by the so-called induction term, the second one in the circuit equation
\be
   \oint \E\cdot d{\bf S} = IR + L\frac{dI}{dt}.
   \label{circuit-eq}
\ee
The inductance is $L=\Phi/I$, so 
\be
  L\frac{dI}{dt} = \frac{d\Phi}{dt},
\ee
consistent with (\ref{magneticflux}) and (\ref{circuit-eq}).

To put this into a form useful to us, we use the approximation
\be
  \oint d{\bf l} = D,
\ee
where $D$ is the size of the region.
Then
\be
  \oint \E \cdot d{\bf l} = D\E, 
  \label{EandD}
\ee
and
\be
   \frac{d\Phi}{dt} = D^2 \partial_t\B.
\ee
So 
\be
  \E = D\partial_t \B.
  \label{EandB}
\ee
We remind the reader that this is the field required to overcome the effect of a growing magnetic field, just as $\eta\J = (\eta/4\pi)\nabla\times\B$ is the field required to overcome resistance.
Since the latter vanishes in our case, (\ref{EandB}) is our equation for $\E$.

Using it, we show that
\be
  \E\cdot \B = D\B\cdot\partial_t\B = D \partial_t\left(\frac{1}{2}\B^2\right).
\ee
Since
\be
  \E\cdot\B = -\frac{1}{2} \partial_t(\A\cdot\B),
\ee
we find that
\be
  \partial_t(\A\cdot \B + D\B^2)=0,
\ee
showing that it takes energy to create helicity.
This is physically reasonable, given that the helicity is a measure of the topology of the magnetic configuration.

\section{Creation of Helical Fields}

In this section we use the above relations to reconsider the evolution of the magnetic field during the axion's rolling period, including effects of the expanding universe.
Let us consider the electromagnetic equations of motion, but this time in curved spacetime, where the Lagrange density is
\begin{equation}
  {\cal L} = -\sqrt{g}\left( {1 \over 4}F_{\mu\nu}F^{\mu\nu} +
  {\g \over 4}\varphi 
   F_{\mu\nu}{\widetilde F}^{\mu\nu}
  \right) \ .
  \label{lagrangian} 
\end{equation}
The equations of motion for $F_{\mu\nu}$ are
\begin{equation}
  \nabla_\mu F^{\mu\nu} = -\g(\nabla_\mu
  \varphi){\widetilde F}^{\mu\nu} \ ,\label{maxwell} 
\end{equation}
along with the Bianchi identity
\begin{equation}
  \nabla_\mu {\widetilde F}^{\mu\nu} = 0 \ .\label{bianchi} 
\end{equation}

We now specialize to the Robertson-Walker metric (\ref{rwmetric}), and set $\varphi = \varphi(t)$.
It will be most convenient to use conformal time  $\t = \int d\t = \int R^{-1}dt$, so that the metric is
\be
  ds^2 = R^2(\t) (-d\t^{2} + d\x^2)\ .
  \label{conformal}
\ee
For reference we note that the scale factor during the radiation-dominated phase behaves as 
\be
  R(t) = R_{*} \left(t\over{ t_*}\right)^{1/2}\ ,
  \label{raddom}
\ee
where $t_*$ is some reference time.
The conformal time therefore is
\be
  \hat{t} = 2{{t_{*}^{1/2}}\over{R_{*}}} t^{1/2}\ ,
\ee
so we can write the scale factor as
\be
  R(\hat t) = \frac{R_*^2}{2t_*}\hat{t}.
\ee

In order to gain intuition, we consider a simplified approximate version of the evolution of $\varphi$.
Before the QCD phase transition, the axion has no potential, so $\varphi$ stays at a constant value $\varphi_0$, and no helicity is generated.
Let $\hat t_1$ be the conformal time when the axion begins to evolve.
We will approximate its evolution by a linear motion from $\varphi=\varphi_0$ to $\varphi=0$, reaching $0$ at a conformal time $\hat t_2$.
After that the field undergoes damped oscillations, but their net effect should be small, so we approximate $\varphi'=0$ after $t_2$.

We therefore have three periods: $\mathcal{A} = (\hat t<\widehat  t_1)$, in which the field is constant; $\mathcal{B} = (\hat t_1 \leq \hat t \leq \hat t_2 = \hat t_1+\Delta \hat t)$, in which $\varphi$ evolves at a rate
\be
  \varphi' = - \frac{\varphi_0}{\Delta \hat t},
\ee
and $\mathcal{C} = (\hat t>\hat t_2)$, when the field is once again constant.
From the numerical results of the last section, for the parameters chosen we have $t_1= 1.0 \times 10^{-7}$\,s and $t_2 = 1.6\times 10^{-7}$\,s.
For the conversion to conformal time, let us choose
\be
  t_* = t_1, \quad R_* = R(t_1) = 1.
\ee
Then from (\ref{raddom}),
\be
  R(t_2) = \left(t_2\over{ t_1}\right)^{1/2} = 1.26.
  \label{R2}
\ee
We then have
\bea
  \hat t_1 = 2t_1 = 2.0\times 10^{-7}\, \mathrm s, \nonumber \\
  \hat t_2 = 2\sqrt{t_1 t_2} = 2.5\times 10^{-7}\, \mathrm s, \nonumber \\
  \Delta \hat t= \hat t_2 - \hat t_2  = 5\times 10^{-8}\, \mathrm s.
  \label{bart}
 \eea
Thus with $\varphi_0 = 1$ we have, in period $\mathcal{B}$,
\be
  \varphi' = -2\times 10^7\, \mathrm{s}^{-1}.
  \label{phiprime}
\ee
This is a crude approximation, but will serve for obtaining an order-of-magnitude estimate of the effect and its parametric dependence on the relevant quantities.

The equations of motion are more transparent if we define the {\bf E} 
and {\bf B} fields by
\bea
  F_{\mu\nu} = R^2 
  \begin{pmatrix}
  0 & -E_x & -E_y & -E_z \cr
  E_x & 0 & B_z & -B_y \cr E_y & -B_z & 0 & B_x \cr
  E_z & B_y & -B_x & 0 
  \end{pmatrix}
  , \nonumber\\
   F^{\mu\nu} = R^{-2}
   \begin{pmatrix}
    0 & E_x & E_y & E_z \cr
        -E_x & 0 & B_z & -B_y \cr
        -E_y & -B_z & 0 & B_x \cr
        -E_z & B_y & -B_x & 0 
        \end{pmatrix} ,\label{e-and-b} 
\eea
or equivalently
$F_{i0} = R^2 E_i$, $F_{ij} = R^2 \epsilon_{ijk} B_k$.  
For reference we can also include the dual tensor,
\bea
  \widetilde{F}_{\mu\nu} = R^2 
  \begin{pmatrix}
 0 & -B_x & -B_y & -B_z \cr
  B_x & 0 & -E_z & E_y \cr 
  B_y & E_z & 0 & -E_x \cr
  B_z & -E_y & E_x & 0 
  \end{pmatrix}
  ,\nonumber \\
   \widetilde{F}^{\mu\nu} = R^{-2}
   \begin{pmatrix}
    0 & B_x & B_y & B_z \cr
        -B_x & 0 & -E_z & E_y \cr
        -B_y & E_z & 0 & -E_x \cr
        -B_z & -E_y & E_x & 0 
        \end{pmatrix} ,
  \eea
from which we verify that $F_\mn \widetilde{F}^\mn = -4 \E\cdot \B$.

Then (\ref{maxwell}) becomes
\begin{equation} 
  {\partial \over \partial \t}(R^2{\bf E}) - \nabla \times (R^2{\bf 
   B}) = -\g
  {\partial \varphi \over \partial \t}R^2{\bf B} - 4\pi R^4 \J\ , \label{4.7} 
\end{equation}
with
\begin{equation}
  \nabla \cdot {\bf E} = 0 \ .\label{4.8} 
\end{equation} 
Here we have included an appropriately-scaled form of the current $\J$, following \cite{PhysRevD.58.083502}.
The Bianchi identity becomes
\begin{equation}
  {\partial \over \partial \t}(R^2{\bf B}) + \nabla \times 
  (R^2 {\bf E}) = 0 \ ,\label{4.9} 
\end{equation}
with
\begin{equation} 
  \nabla \cdot {\bf B} = 0 \ . \label{4.10} 
\end{equation}
In these expressions, $\nabla$ represents the usual gradient operator, expressed in comoving coordinates.

It is helpful to rescale the electromagnetic field via:
\be
  \F = R^2\B\ ,\quad \G = R^2\E\ .
  \label{rescaled}
\ee
Then we can write (\ref{4.7}) and (\ref{4.9}) as
\be
  {\partial \over \partial \t}\G - \nabla \times \F = -\g{\partial \varphi \over \partial \t}\F - 4\pi \widehat\J
    \ , \label{eq11}
\ee
and
\be
  {\partial \over \partial \t}\F + \nabla \times \G,= 0 \ ,\label{faraday}
\ee
where we have defined $\widehat\J = R^4 \J$.
We see that the conformal-time equations for the rescaled variables take the form of the conventional Maxwell equations (with our additional pseudoscalar interaction).

In periods $\mathcal{A}$ and $\mathcal{C}$, $\varphi'=0$ (where a prime indicates $\partial/\partial\t$), so standard MHD applies.
According to (\ref{standardE}), with $\eta=0$ we then have $\widehat\E= \v\times\widehat\B$, so 
\be
  \partial_{\t}\widehat\B = - \nabla\times(\v\times\widehat\B).
\ee
This is the equation that Alfv\'en used to show that the magnetic lines of force are frozen into the plasma, and therefore move with it. 
Following FC, we assume this is the case in period $\mathcal{C}$, after magnetic helicity is created in period $\mathcal{B}$.
In ideal MHD $\partial_{\t}\widehat\E$ is neglected on the grounds that it is needed only to discuss electromagnetic waves \cite{2017mcp..book.....T}.
Then ccording to (\ref{eq11}), $4\pi\widehat\J = \nabla\times\widehat\B$, Ampere's Law.
Once we enter period $\mathcal{B}$, $\varphi'\neq 0$, so the penultimate term in (\ref{eq11}) no longer vanishes.

Our task is to find the solution to (\ref{eq11}) in period $\mathcal{B}$. 
To simplify (\ref{eq11}) we set
\be
  4\pi\widehat\J = \nabla \times \widehat\B,
  \label{jdivb}
\ee
as is done in ordinary MHD with $\varphi'=0$.
This introduces a slight problem.
Consider the divergence of (\ref{eq11}).
It yields
\be
  \partial_{\t}(\nabla \cdot\widehat\E) + \nabla\cdot(4\pi\widehat\J) = 0,
\ee
which, because $\nabla\cdot\widehat\E = 4\pi \hat\rho$ and $\nabla\cdot\widehat\J = -\hat\rho$, represents charge conservation.
If on the other hand (\ref{jdivb}) is applied, the divergence yields
\be
  \partial_\t \hat\rho = 0.
\ee
While this is not always precisely true, it is a good approximation in the situation we are considering.
The reason is that any deviation of $\hat\rho$ from zero results in plasma oscillations at the frequency
\be
  \omega_\mathrm{p} = \sqrt{\frac{4 \pi n_{\mathrm{e}} e^2}{m_e}}.
\ee
This is far greater than the inverse timescales under consideration here.
We therefore continue with (\ref{jdivb}).

This leaves us with the following equation of motion for the electric field:
\be
\partial_{t} \widehat\E=-\g\varphi' \widehat\B.
\label{partialE}
\ee
The solution to (\ref{partialE}) depends upon the behavior of $\widehat\B$, which is determined by the Bianchi identity (Faraday's Law) (\ref{faraday}).
Taking the time derivative and using (\ref{partialE}), we have
\bea
\partial_{\t}^{2} \widehat\B &=&-\nabla \times \partial_{t} \widehat\E \nonumber \\
 &=&\g \varphi' \nabla \times \widehat\B.
\label{ddotB}
\eea
Note that, given our assumption $\nabla\times\widehat\B = 4\pi\widehat\J$, the actual value of the resistivity has dropped out, and creation of electromagnetic fields can occur even if the plasma is highly conductive.

Considering the spatial Fourier transform,
\be
  \widehat\B_k(\t) = \int e^{i \k\cdot\x} \widehat\B(\x,\t)\, d^3x,
\ee 
(\ref{ddotB}) implies
\be
  \partial_{\t}^{2} \widehat\B_k = -i\g\varphi' \k\times\widehat\B_k.
\ee
To consider the evolution of a single mode, let $\k = k\hat{x}$, and define
\be
  \widehat B_\pm = \widehat B_y \pm i \widehat B_z.
\ee
Then the equation of motion for $\widehat B_\pm$ is
\be
  \partial_{\t}^{2}\widehat B_\pm \pm \g\varphi' k \widehat B_\pm = 0.
\ee

We have adopted an approximation according to which, during period $\mathcal{B}$, we have $\varphi' = - 2\times 10^7 \mathrm{s}^{-1} =$ constant.
The solution in this period is given by:
\be
  \widehat B_\pm(\hat t) = \widehat B_\pm(\hat t_1) \exp{n(\hat t - \hat t_1}),
  \label{Bsolution}
\ee
where  the growth rate is
\be
  n = \pm \sqrt{|\g\varphi'_0 k|}.
  \label{growthrate}
\ee
At the end of period $\mathcal{B}$, we have
\be
\widehat B_\pm(\hat t_2) = e^\beta\widehat B_\pm(\hat t_1) ,
\ee
where the exponent is 
\be
  \beta = n\Delta\hat t  = \pm |\g(\Delta\varphi) k (\Delta\hat t)|^{1/2}.
  \label{beta}
\ee
We see that there is potential exponential growth, whose magnitude will increase at shorter wavelengths (larger $k$).
To convert back to the conventional magnetic field, we use (\ref{R2}) and (\ref{rescaled}) to get
\be
  \B_\pm(t_2) = \left(\frac{R(t_1)}{R(t_2)}\right)^2 e^\beta \B_\pm(t_1) = 0.79 e^\beta \B_\pm(t_1).
\ee

\section{Helicity}

Now let us consider the helicity.
Our results of the last section imply that the expansion of the universe only introduces an overall order-of-magnitude multiplicative factor into our results, essentially because period $\mathcal{B}$ doesn't last very long.
So for simplicity here we will return to using flat-spacetime notation.
Because the magnetic field freezes in once we go from period $\mathcal{B}$ to $\mathcal{C}$, the expansion of the universe can be accounted for by letting $\B$ scale as $R^{-2}$.
In addition, there is a scaling of the correlation length due to the inverse cascade.

The time derivative of the helicity density is
\be
  \partial_{t}(\A\cdot\B) = -2 \E \cdot \B,
\ee
where we are ignoring a total derivative, since we ultimately care about the spatial integral of this quantity.
From (\ref{partialE}) we have
\be
  \B = -\frac{1}{\g\varphi'}\partial_{t} \E,
  \label{BofE}
\ee
so
\be
  \partial_{t}(\A\cdot \B) = \frac{2}{\g\varphi'} \E \cdot \partial_{t}\E = \frac{1}{\g\varphi'}\partial_{t} (\E)^2.
\ee
We therefore have
\be
  \Delta(\A \cdot \B) = \frac{1}{\g\varphi'}\Delta\E^2.
\ee
In period $\mathcal{B}$, where $\varphi'$ is constant, from (\ref{Bsolution}) and (\ref{BofE}) we have
\bea
  \E &=& \g\varphi' \int_{t_1}^{t_2} \B \, dt \nonumber \\
  &=& -\g\varphi'\B(t_1) \int_{t_1}^{t_2}  e^{nt}\, dt \nonumber \\
  &=& \frac{\g\varphi'}{n} \B(t_1) (e^{\beta} - 1),
\eea
where $\beta = n\Delta t$ is given by (\ref{beta}) (and we approximate $\Delta \hat t$ as $\Delta t$).
Therefore,
\be
  \Delta(\A \cdot \B) = \frac{(\g\varphi')^2}{n^2} \frac{1}{\g\varphi'} \B^2(t_1)
  (e^{2\beta} - 2 e^{\beta} + 1).
\ee
And since from (\ref{growthrate}) $n^2 = gk\g\varphi'$,
\be
  \Delta (\A\cdot \B) = \frac{1}{gk}\B^2(t_1)f(\beta),
\ee
where
\be
  f(\beta) = e^{2\beta} - 2 e^{\beta} + 1.
  \label{fbeta}
\ee
We calculate some representative values of $\beta$ and $f(\beta)$ in Table 1 below.

The importance of helicity comes from the claim from FC that a maximally-helical cosmic magnetic field will undergo an inverse cascade to longer wavelengths.
This is crucial, as the small horizon size in the early universe means that any primordial field will have a coherence length no longer than the Hubble radius at that time, which is smaller (even after the universe expands) than the scales of current astrophysical interest.
For a maximally helical field, the current correlation length $L_0$ can be expressed as
\be
  L_0 = S(t_0)\left(\frac{R_0}{R_*}\right)L_*,
  \label{L0}
\ee
where $L_*$ is the initial coherence length, the factor $R_0/R_*$ reflects the Hubble expansion, and $S(t)\geq 1$ is a ``cascade factor" from the inverse cascade of $H_M$.

FC showed that there is a scaling relation according to which $t \rightarrow S t$, $\Delta t \rightarrow S \Delta t$, and $\B \rightarrow S^{-1 / 2}\B$. 
They argued that, in the radiation-dominated era, this led to the cascade factor being given by
\be
  S = \left(\frac{t}{t_0}\right)^{-1/3},
  \label{cascadefactor}
\ee
where $t_0 = 4\times 10^{17}\,\mathrm{s}$ is the current age of the universe.
Here we will assume that this same functional form translates to the matter- and vacuum-dominated eras as well.
This is reasonable as long as we express $S$ as a function of proper time $t$ rather than as a function of the scale factor.
Any deviation should be sub-dominant, since the total increase in the scale factor happens mostly in the radiation-dominated era.

\section{Numerical Estimates}

We now plug in numbers to gauge the effectiveness of this mechanism in generating primordial magnetic fields.

In Section~\ref{pseudoscalar} we saw that the epoch $\mathcal{B}$ of axion rolling is at $t_{*} \approx 1.5\times 10^{-7}$~sec, corresponding to $T_{*} = 2.0$~GeV.
The scale factor is related to the temperature and the effective number of relativistic degrees of freedom $g_{*S}$ by $R \propto g_{*S}^{-1/3} T^{-1}$, where $g_{*S}$ is approximately 80 during period $\mathcal{B}$ and 3.36 today.
This gives
\be
  \frac{R_{*}}{R_0} = \frac{T_0}{T_{*}}\left(\frac{g_{*S}(t_0)}{g_{*S}(2\,\mathrm{GeV})}\right)^{1/3}
  = 4.0\times 10^{-14},
  \label{scalefactor}
\ee
where $R_0$ is the scale factor today.
Meanwhile the cascade factor (\ref{cascadefactor}) between period $\mathcal{B}$ and today is
\be
  S = 1.4 \times 10^8.
\ee

Consider the amplification of the field during period $\mathcal{B}$, given by $e^\beta$, with $\beta=|\g(\Delta\varphi) k (\Delta\hat t)|^{1/2}$ from (\ref{beta}).
To get numerical estimates, we set $\Delta\varphi = 1.0$, a reasonable value for a random number between $0$ and $\pi$.
The coupling constant $\g$, for typical particle-physics parameters, takes on values of order $10^{-2}$, but values as large as $\g \approx 10$ are compatible with current data \cite{Workman:2022ynf}.
Choosing to be somewhat optimistic, we set $\g=2$.
Finally we have $\Delta\hat t = 6\times 10^{-8} \,\mathrm{s} = 1800\,\mathrm{cm}$.

We need to know what values of the wave number $k$ are of physical relevance.
Any physical scale $\lambda = 1/k$ of the helical magnetic field in the axion epoch will be stretched both because of the scale factor and the cascade factor, so a given scale of interest in the current universe $L_0$ corresponds to 
\be
  L_0 = S\frac{R_{*}}{R_0}k^{-1}.
\ee
That gives
\be
  k = \frac{1.1}{L_0 (\mathrm{kpc})}\, \mathrm{cm}^{-1}.
\ee
The Hubble radius $ct$ at $T=2\,$GeV is 
\be
  D_H = 1/(2H)= 2000\,\mathrm{cm},
\ee
so causally relevant scales are those for which $L_0 \leq 2000\,$kpc.
Putting it all together, we have
\be
  \beta = \frac{44.5}{L_0 [\mathrm{kpc}]^{1/2}} .
\ee

We have indicated some representative values in Table 1, including the helicity growth factor $f(\beta)$ defined in (\ref{fbeta}).
We conclude that the amplification of the field is significant only if $L_0 < 100\,$kpc (although the growth could be larger with larger values of $g_\varphi$ or $\Delta\varphi$).
There is substantial growth at smaller scales, but any initial field that could be amplified might also be smaller on sub-Hubble scales.
\begin{table}[h]
\caption{}
\begin{tabular}{cccc}
\\
 $L_0$ (kpc) & $\beta$ & $e^\beta$ & $f(\beta)$ \\
\hline \\
3 & 36 & ~~$5.9\times 10^{15}$~~ & $3.5 \times 10^{31}$ \\
10 & 20 & ~~$4.3\times 10^{8}$~~ & $1.9 \times 10^{17}$ \\
30 & 11 & $9.7\times 10^4$ & $9.4 \times 10^9$ \\
100 & 6.3 & $5.4 \times 10^2$ & $2.9\times 10^5$ \\
 300& 3.6 & 38 & $1.4\times 10^3$ \\
 1000 & ~~2.0~~ & ~~7.3~~ & ~~40~~
\end{tabular}
\end{table}

The mechanism we have proposed here takes a pre-existing field and amplifies it due to the evolution of the axion field.
We have not proposed an origin for the pre-existing field, but we can figure out what its magnitude should be.
Once we make the transition from period $\mathcal{B}$ to $\mathcal{C}$, the magnetic field is frozen into the expanding plasma, so that
\be
  B_0 = \left(\frac{R_*}{R_0}\right)^2 B_*,
  \label{Bscaling}
\ee
where $B_*$ is the field magnitude at its creation, when the scale factor is $R_*$.
The value of $B_0$ is not known directly, but it can be inferred from \cite{2002A&A...387..383D}, who found that the observed values of $B$ in clusters of galaxies today fits the hypothesis that they are primordial fields that were later accreted by the clusters.
Their hypothesis requires that the field at $z=0.7$ is $B_0=1.3$\,nG. 
Let us adopt this as an approximate value of the current intergalactic magnetic field at correlation length of $L_0 = 30\,$kpc.
This corresponds to a scale of 
\be
  L_* = S^{-1} \left(\frac{R_*}{R_0}\right)30\,\mathrm{kpc} = 27\,\mathrm{cm},
\ee 
about $10^{-2}$ of the Hubble scale during period $\mathcal{B}$.
It follows that the value $B_*$ at the time of interaction is
\be
  B_* = \left(\frac{R_*}{R_0}\right)^{-2}B_0 = 8.1 \times 10^{17}\,\mathrm{G}.
\ee
Tthe energy density of the magnetic field is then
\be
  \rho_B = \frac{B_*^2}{8\pi} = 2.7\times 10^{34}\,\mathrm{erg\,cm}^{-3}.
\ee
We can compare this to the total energy density at $T_*=2\,$GeV,
\be
  \rho_*  = \frac{\pi^2}{30}g_*(T)T^4 = 8.9 \times 10^{40}\,\mathrm{erg\, cm}^{-3},
\ee
so the fraction of the energy density in the magnetic field is
\be
  f_M = 3.0 \times 10^{-7}.
  \label{fraction}
\ee
This seems like a plausible target for a pre-existing magnetic field that could be amplified, but this issue merits further investigation.

\section{Conclusion}

This work is an attempt to follow up on a suggestion by Field and Carroll \cite{Field:1998hi} that a primordial magnetic field must be helical if it is to have intergalactic size today, as proposed by Dolag et al \cite{2002A&A...387..383D}. 
FC suggested that such a helical field could be produced by interaction between a magnetic field and a pseudoscalar coupled via $\varphi F_{\mu\nu}\tilde{F}^{\mu\nu}$, of which the axion is an example. 
Such a Lagrangian leads to the modified Maxwell equation
\be
  \partial_t\E + 4\pi \J - \nabla\times\B + \g\dot\varphi\B = 0
  \label{con1}
\ee 
when $\varphi$ depends only on $t$.
Magnetic helicity $H=\int \A\cdot\B \, d^3x$ is created by
\be
  \partial_t(\A\cdot\B) = - 2\E\cdot\B = \frac{1}{2}F_{\mu\nu}\tilde{F}^{\mu\nu},
\ee
so an electric field must be present; such a field is also required if a magnetic field is to be generated according to
\be
  \partial_t \B = -\nabla \times\E,
\ee
but helicity can be created even if $\B$ is already present.

To solve (\ref{con1}) we chose to simplify it by mutually canceling the terms $4\pi \J$ and $\nabla\times\B$, which together form Ampere's Law.
We have offered a justification for why this simplification is appropriate in the conditions we study.

Under this assumption, in the presence of an evolving pseudoscalar field, we have shown that the helical component of the magnetic field can be substantially amplified.
We have explored the prospects for such amplification if the pseudoscalar is the axion, evolving in the era just before the QCD phase transition, when $T\sim 2\,$GeV.
We conclude that any field that is coherent over slightly-sub-Hubble-radius scales could undergo substantial growth.
One could imagine a precursor field arising from the electroweak phase transition, with $T_\mathrm{EW}\sim 200\,$GeV being just two orders of magnitude higher than the axion era, but we did not pursue any specific mechanisms here. 
If any mechanism provided such a field, axion evolution could boost it to astrophysically interesting length scales and magnitudes.

Our model predicts that there are nanogauss fields at multiple-kiloparsec scales between the galaxies. 
Given the results of this paper, it is important to search for the existence of intergalactic magnetic fields, and to work to measure their coherence length, as well as their helicity if possible.
Such measurements could give important clues as to whether the fields originated in the early universe.

\bibliographystyle{utphys}
\bibliography{bfields}
\end{document}